\def\NPA{{Nucl. Phys.} {\bf A}}
\def\NPB{{Nucl. Phys.} {\bf B}}
\def\PLB{{Phys. Lett.} B}

\def\PRD{{Phys. Rev.} D}

\documentclass[
    ,final            % use final for the camera ready runs
  ]
  {aipproc}

\layoutstyle{6x9}
\usepackage{bm}% bold math
\begin{document}

\title{Infrared Features of Lattice Landau Gauge QCD\\ and the Gribov Copy Problem}

\author{Sadataka Furui}{
  address={School of Science and Engineering, Teikyo University, Utsunomiya 320-8551,Japan}
}

\author{Hideo Nakajima}{
  address={Department of Information science, Utsunomiya University, Utsunomiya 320-8585,Japan }
}

%\author{<author3>}{
%  address={<common address for author2 and author3>}
%  ,altaddress={<author1 address>} % additional visiting address
%}

\begin{abstract}
Infrared features of gluon propagator, ghost propagator, QCD running coupling
and the Kugo-Ojima parameter in lattice Landau gauge QCD are presented. The
framework of PMS analysis suggests that there appear infrared, intermediate and ultraviolet regions specified by $\Lambda_{\overline{MS}}$, $\beta_0$ and $\beta_1$.  The propagators and the running coupling of $q>1GeV$ are fitted by
 the $\widetilde{MOM}$ scheme using the factorization scale as the scale parameter. The running coupling data of $q<14GeV$ are fitted by the contour improved
perturbation method. The Gribov copy problem is studied in $SU(2)$, $\beta=2.2$ samples by comparing the data gauge fixed by the parallel tempering method
and the data gauge fixed by the straightforward gauge fixing. The Gribov noise effect turned out to be about 6\%.
\end{abstract}

\maketitle

%%%%%%%%%%%%%%%%%%%%%%%%%%%%%%%%%%%%%%%%%%%%
%% MAINMATTER
%%%%%%%%%%%%%%%%%%%%%%%%%%%%%%%%%%%%%%%%%%%%

\section{Introduction}

  In Landau gauge QCD, it is necessary to restrict the gauge configuration
expressed by the link variable $U_{x,\mu}$ 
to  $\Omega_L=\{U|-\partial { D(U)}\ge 0\ ,\ \partial A=0\}$ which is called
 Gribov region (local minima)\cite{Gr}, where $D$ is covariant derivative. Zwanziger argued that the uniqueness of the 
gauge field is guaranteed in its subset $\Lambda_L=\{U|\ A=A(U), F_{U}(1)={\rm Min}_gF_{U}(g)\}$, and called  the $\Lambda_L$ fundamental modular(FM) region\cite{Zw}. Here the optimizing function $F_{U}(g)$ is defined in the case of
 $U-$linear ($A_{x,\mu}=\displaystyle{1\over 2}(U_{x,\mu}-U_{x,\mu}^{\dag})|_{trless\ p.}$) and $\log U$ ($U_{x,\mu}=e^{A_{x,\mu}})$ as, 
$F_U(g)=\sum_{x,\mu}\left (1- {1\over 3}{\rm Re}\ {\rm tr}U^g_{x,\mu}\right),$
and
$F_U(g)=||A^g||^2=\sum_{x,\mu}{\rm tr}
 \left({{A^g}_{x,\mu}}^{\dag}A^g_{x,\mu}\right)$, respectively\cite{NF}.

 In Landau gauge, QCD running coupling can be extracted from three gluon coupling and/or ghost-gluon coupling. In terms of gluon dressing functiuon $Z_A(q^2)$ and ghost dressing function $G(q^2)$, which are  measurable in lattice Landau gauge, renormalization group invariant quatity\cite{SHA} $\alpha_s(q^2)=g^2 G(q^2)^2 Z_A(q^2)/4\pi$ can be measured. 

Colour confinement in infrared QCD is characterized by the Kugo and Ojima 
parameter $u(0)$ which is defined as
\[
\displaystyle{1\over V}
\sum_{x,y} e^{-ip(x-y)}\langle  {\rm tr}\left({\lambda^a}^{\dag}
D_\mu \displaystyle{1\over -\partial D}[A_\nu,\lambda^b] \right)_{xy}\rangle=
(\delta_{\mu\nu}-{p_\mu p_\nu\over p^2})u^{ab}(p^2),
\]
where $u^{ab}(p^2)=\delta^{ab}u(p^2), \ u(0)=-c.$
The sufficient condition of the colour confinement is $u(0)=-1$. 
We measure the running coupling and the Kugo-Ojima parameter on lattice and study the ambiguity due to Gribov copy.

\section{Numerical Results of Lattice Landau gauge QCD }

\subsection{The gluon propagator and the ghost propagator}
\begin{figure}[htb]
\begin{minipage}[b]{0.60\linewidth} 
\begin{center}
\includegraphics{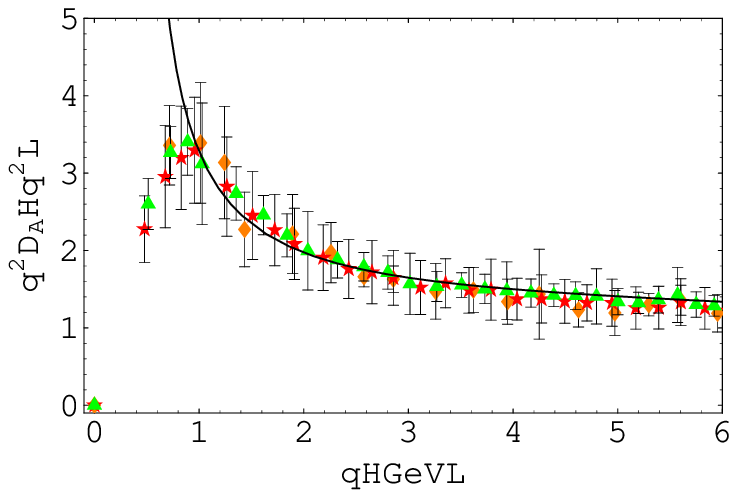}%{gl243248m.eps}
\end{center}
\end{minipage}\hfil
\begin{minipage}[b]{0.60\linewidth} 
\begin{center}
\includegraphics{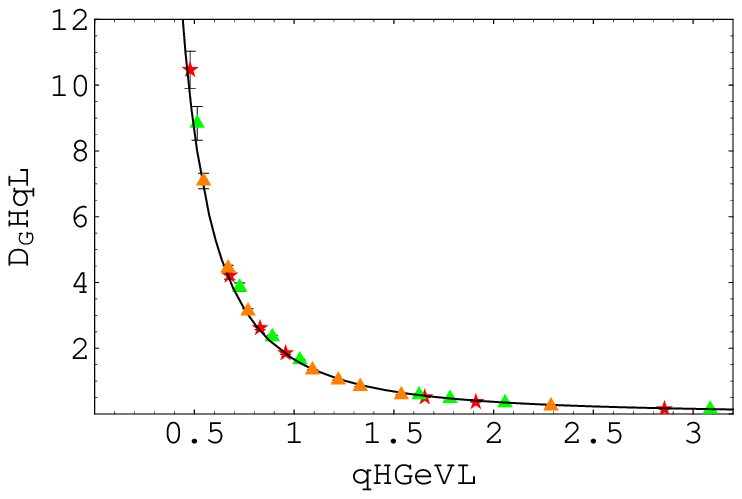}%{gh243248.eps}
\end{center}
\caption{(a)The gluon dressing function as the function of the momentum $q(GeV)$. $\beta=6.0$, $24^4$(triangle), $32^4$(diamond) and $\beta=6.4$, $48^4$(star) in $\log U$ version. The fitted line is that of the $\widetilde{MOM}$ scheme. 
(b)The ghost propagator as the function of the momentum $q(GeV)$. $\beta=6.0$, $24^4, 32^4$ and $\beta=6.4$, $48^4$ in $\log U$ version. The fitted line is that of the $\widetilde{MOM}$ scheme. }\label{gh243248}
\end{minipage}
\end{figure}

The gluon  propagator of colour $SU(n)$ gauge is defined as
\begin{eqnarray}
D_{\mu\nu}(q)&=&{1\over n^2-1}\sum_{x={\bf x},t}e^{-ikx}Tr\langle A_\mu(x)A_\nu(0)^\dagger \rangle \nonumber\\
&=&(\delta_{\mu\nu}-{q_\mu q_\nu\over q^2})D_A(q^2)
\end{eqnarray}
and the gluon dressing function as $Z_A(q^2)=q^2 D_A(q^2)$. Our result is shown in Fig.\ref{gh243248}(a) and is consistent with\cite{adelade, orsay1}.

The ghost propagator, which is the Fourier transform of an expectation value of the inverse Faddeev-Popov operator ${\cal  M}=-\partial D=-\partial^2(1-M)$
\begin{equation}
D_G^{ab}(x,y)=\langle {\rm tr} \langle \lambda^a x|({\cal  M}[U])^{-1}|
\lambda^b y\rangle \rangle
\end{equation}
where the outmost $\langle\rangle$ denotes average over samples $U$, 
is evaluated as follows. We take plane wave for the source
${\bf b}^{[1]}={\bm\lambda}^b e^{iqx}$ and get the solution of Poisson equation ${\bm \phi}^{[1]}=(-\Delta )^{-1}{\bf b}^{[1]}$. 
We calculate iteratively  $\phi^{[i+1]}=M{\bm \phi}^{[i]}({\bf x})$($i=1,\cdots,k-1$).
   The iteration was
continued until $Max_{\bf x}|{\bf \phi}^{[k]}({\bf x})|/Max_{\bf x}|\sum_{i=1}^{k-1}{\bm \phi}^{[i]}({\bf x})|<0.001\sim 0.01$.
The number of iteration $k$ is of the order of 60, in $SU(2)$, $16^4$ lattice,  and of the order of 100 in $SU(3)$. 
We define ${\bm\Phi}^b({\bf x})=\sum_{i=1}^k{\bm \phi}^{[i]}({\bf x})$ and evaluate
$\langle \lambda^a e^{iqx},\Phi^b({\bf x})\rangle$ as the ghost propagator from
colour $b$ to colour $a$. We also used the conjugate gradient(CG) method and observed that the data agree except at the lowest momentum point of $48^4$. Results of CG method are shown at this momentum point.

The gluon and the ghost dressing functions are fitted by using the framework of principle
of minimal sensitivity(PMS)\cite{PMS} or the effective charge method\cite{Gru},
 in which the running coupling $h(q)$ is parametrized as 

\begin{eqnarray}
h(q)&=&y_{\overline{MS}}(q)\{1+y_{\overline{MS}}(q)^2(\bar\beta_2/\beta_0-(\beta_1/\beta_0)^2)\nonumber\\
&+&y_{\overline{MS}}(q)^3\frac{1}{2}(\bar \beta_3/\beta_0-(\beta_1/\beta_0)^3)+\cdots\}.
\end{eqnarray}

The parameter $y_{\overline{MS}}(q)$ can be expressed by the parameter $y$ defined as a solution of 
\begin{equation}\label{scale}
 \beta_0\log\frac{\mu^2}{\Lambda^2}=\frac{1}{y}+\frac{\beta_1}{\beta_0}\log(\beta_0 y)
\end{equation}
where $\Lambda$ characterizes the scale of the system, and the function
\begin{equation}
 k(q^2,y)=\frac{1}{y}+\frac{\beta_1}{\beta_0}\log(\beta_0 y)-\beta_0\log(q^2/\Lambda_{\overline{MS}}^2).
\end{equation}

In PMS, $y$ is treated as a function of $q^2$. However, in this work
we fix the scale by the factorization scale $\mu=1.97GeV$ at which renormalizable quantity can be approximately factorizable into scheme independent and dependent parts\cite{Gru,orsay1}.

In order to be consistent with the $\overline{MS}$ scheme, we define the variable $z$ as 
\begin{eqnarray}
z&=&-e^{(-1-b t/2c)}=-\frac{1}{e}(\frac{q}{\tilde\Lambda_{\overline{MS}}})^{-b/c}e^{iK\pi}\nonumber\\
&=&-Z(q^2)e^{iK\pi}
\end{eqnarray}
where $t=\log (q^2/\tilde\Lambda_{\overline{MS}}^2)$, $c=\beta_1/\beta_0=51/22, b=\beta_0/2=11/2$, $\tilde\Lambda_{\overline{MS}}=(2c/b)^{-c/b}\Lambda_{\overline{MS}}$, $K=-b/2c$\cite{PMS,HoMa}.
When we fix $y$ by the solution of eq.(\ref{scale}) with $\mu=1.97GeV$, ($\widetilde{MOM}$ scheme), we find that in the gluon dressing function the Landau pole at $z=1/e$ 
remains, and another pole appears at $z\sim0.17$. When $y$ is chosen as $q^2$
dependent, the three regions $0<z<0.17$,  $0.17<z<1/e$ and $1/e<z$ can be continuously connected\cite{vAc}, but there is a subtle problem of PMS in low energy\cite{BroLu} and leave the problem to bridge the three regions to the future.

The gluon dressing function $q^2D_A(q^2)$ and the ghost propagator $D_G(q^2)$ in $\widetilde{MOM}$ scheme are plotted in Fig.\ref{gh243248}(a) and (b), respectively. They are singular at $q=\tilde\Lambda_{\overline{MS}}\simeq 0.25GeV$ which should be washed away by the non-perturbative effects. 

\subsection{The QCD running coupling and the Kugo-Ojima parameter}
The QCD running coupling  $\alpha_s(q)$ turned out to have a peak of 
the order of 1 at $q\sim 0.5GeV$ and decreases to a finite value at $q=0$.
We fitted the data by the contour improved perturbation series, which is
a way to make a resummation of the series of coupling constant.

The effective running coupling in
the $\overline{MS}$ scheme is expressed by
the series of coupling constant $h^{(n)}$ as\cite{vAc}.
\begin{equation}
{\cal R}^n=h^{(n)}(1+A_1h^{(n)}+A_2h^{(n)2}+\cdots+A_nh^{(n)n})\label{req}
\end{equation}

\begin{figure}[htb]
\begin{minipage}[b]{0.60\linewidth} 
\begin{center}
\includegraphics{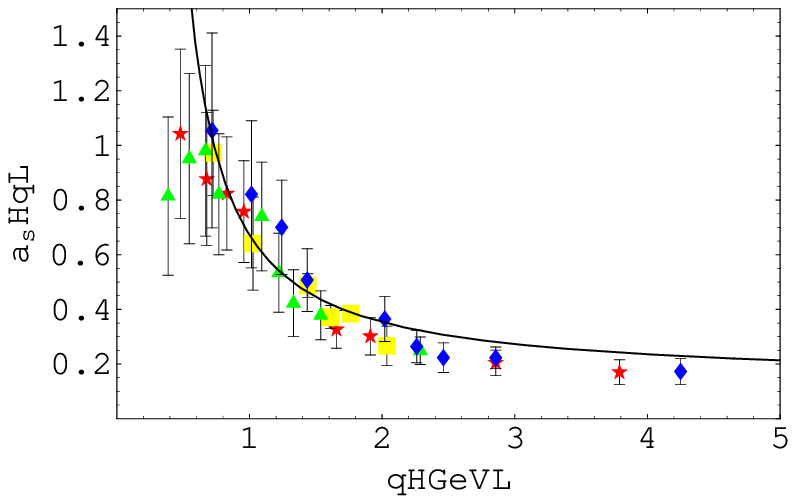}%{alp2432ab48.eps}
\end{center}
\end{minipage}
\hfill
\begin{minipage}[b]{0.60\linewidth}
\begin{center}
\includegraphics{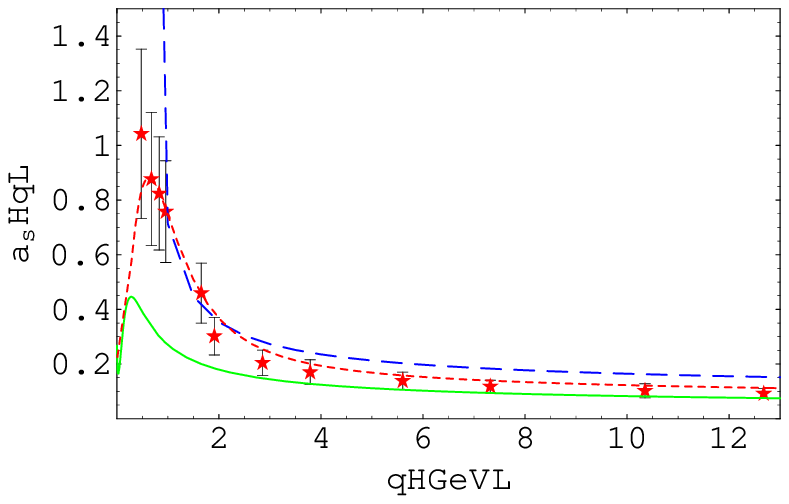}%{alp_48_3.eps}
\end{center}
\caption{(a) The running coupling $\alpha_s(q)$ of $\beta=6.0$, $24^4$(box), 
$32^4$(triangle), $\beta=6.4$, $32^4$(diamond) and $48^4$(star) 
as a function of momentum $q(GeV)$ and the result of the 
 $\widetilde{MOM}$ scheme. 
(b) The running coupling $\alpha_s(q)$ as a function of momentum $q(GeV)$ of $\beta=6.4, 48^4$ lattice. The solid line is the result of ${\cal R}^2$ using $\Lambda_{\overline{MS}}=237MeV$. Dotted line is 
the result of $e^{70/6\beta_0}\Lambda_{\overline{MS}}$ and including half of $A_2$.
Dashed line is the result of $\widetilde{MOM}$ scheme. 
}\label{alppert1}
\end{minipage}
\end{figure}

The result of $\widetilde{MOM}$ scheme using $y=0.01594$ is shown by the solid line in Fig. \ref{alppert1}(a). The lattice data of $24^4, 32^4$ and $48^4$ and the $\widetilde{MOM}$ scheme agree in $0.5GeV<q<2GeV$ but slightly overestimates
in $q>2GeV$.

In contour improved perturbation method, physical quantities ${\cal R}$ are expressed in a series
physical quantities ${\cal R}$ are expressed in a series
\begin{equation}\label{pert}
{\cal R}(q^2)={\cal B}_1(q^2)+\sum_{n=1}^\infty A_n{\cal B}_{n+1}(q^2)
\end{equation}
\begin{equation}
{\cal B}_n=\frac{1}{2\pi}\int_{-\pi}^\pi (\frac{-1}{c(1+W(Z(q^2)e^{iK\theta})})^n d\theta 
\end{equation}
Terms in the series (\ref{pert}) have alternating sign and $A_3$ is not known. By choosing half of $A_2$ in the series, we can obtain a good fit of the data(Fig.\ref{alppert1}(b)).

The Kugo-Ojima parameter depends slightly on definitions, $U-$linear or $\log U$ and
$c$ becomes larger as the lattice size becomes large. However the
value saturates at about 0.8. 
\begin{table}[htb]
\begin{tabular}{c|c|ccc|ccc}
 $\beta$&$L$ &$c_1$ & $e_1/d$ & $h_1$ & $c_2$ & $e_2/d$ & $h_2$ \\
\hline
6.0 &16&  0.576(79) &   0.860(1) & -0.28 & 0.628(94)& 0.943(1) & -0.32\\
6.0 &24&  0.695(63)  &  0.861(1) & -0.17 & 0.774(76)& 0.944(1) & -0.17\\
6.0 &32&  0.706(39)  &  0.862(1) & -0.15 & 0.777(46)& 0.944(1) & -0.16\\
\hline
6.4 &32& 0.650(39) & 0.883(1) & -0.23 & 0.700(42)& 0.953(1) & -0.25\\
6.4 &48&           &          &       & 0.793(61)& 0.982(1) & -0.19\\
\end{tabular}
\caption{The Kugo-Ojima parameter $c$, trace $e/d$ and horizon function deviation $h$  in  $U-$linear and $\log U$  version. $\beta=6.0$ and $6.4$.}\label{kugotab}
\end{table}

\section{The Gribov copy problem}
We applied the parallel tempering(PT) method, which is used in finding global minimum in spin glass systems, to perform the FM gauge fixing of $SU(2)$ $\beta=2.2, 16^4$ lattice configuration and compared
with ensemble of 1st copy(Gribov copy obtained by the straightforward gauge fixing)\cite{lat03}. The Kugo-Ojima parameter $c$ of PT becomes smaller than 1st copy by about $4\%$. Gribov noise in  the ghost propagator is similar to Cucchieri's noise\cite{cuc} at $\beta=1.6$, i.e. the infrared singularity of PT is weaker than 1st copy by about $6\%$. The SU(2) gluon propagator, ghost propagator and running coupling in $q\geq 1GeV$ region are consistent with results of\cite{BCLM}. The running coupling in our simulation has a peak at around $q=1GeV$ and is suppressed in the infrared.

We aimed to detect in the lattice dynamics, the signal of Kugo-Ojima confinement
 criterion derived in the continuum theory, formulated in use of Faddeev-Popov Lagrangian and BRST symmetry. We also noted Zwanziger's horizon condition, based on the lattice formulation, coincides  with Kugo-Ojima criterion\cite{Zw,NF}. However, our present data  are not satisfactory to prove or disprove the confinement criterion. The colour off-diagonal antisymmetric part of ghost propagator\cite{dudal,KoShi} does not appear in Landau gauge, and the off-diagonal symmetric part has vanishing statistical average but has fluctuation proportional to $(qa)^{-4}$.

%%%%%%%%%%%%%%%%%%%%%%%%%%%%%%%%%%%%%%%%%%%%%%%%
%% BACKMATTER
%%%%%%%%%%%%%%%%%%%%%%%%%%%%%%%%%%%%%%%%%%%%%%%%

\begin{theacknowledgments}
We are grateful to Daniel Zwanziger for enlightenning discussion. S.F. thanks Kei-Ichi Kondo, Kurt Langfeld, Karel van Acoleyen and David Dudal for valuable information.
This work is supported by the KEK supercomputing project No.03-94.
\end{theacknowledgments}

%%%%%%%%%%%%%%%%%%%%%%%%%%%%%%%%%%%%%%%%%%%%%%%%
%% You may have to change the BibTeX style below, depending on your
%% setup or preferences.
%%
%% If the bibliography is produced without BibTeX comment out the
%% following lines and see the aipguide.pdf for further information.
%%
%% For The AIP proceedings layouts use either
%%%%%%%%%%%%%%%%%%%%%%%%%%%%%%%%%%%%%%%%%%%%

%\bibliographystyle{aipproc}   % if natbib is available
\bibliographystyle{aipprocl} % if natbib is missing

%%%%%%%%%%%%%%%%%%%%%%%%%%%%%%%%%%%%%%%%%%%
%% You probably want to use your own bibtex database here
%%%%%%%%%%%%%%%%%%%%%%%%%%%%%%%%%%%%%%%%%%%
%\bibliography{sample}

\begin{thebibliography}{99}
\bibitem{KO} T. Kugo and I. Ojima, {Prog. Theor. Phys. Supp.} {\bf 66}, 1 (1979).

\bibitem{Gr} V.N. Gribov, {\NPB} {\bf 139}{1}{(1978)}.

\bibitem{Zw} D. Zwanziger, {\NPB} {\bf 364} ,{127} {(1991)}, idem B
{\bf 412}, {657} (1994).

\bibitem{NF} H.Nakajima and S. Furui, {\NPB} (Proc Suppl.){\bf 63A-C},{635, 865}(1999), %hep-lat/9809080,9809081; 
 {\NPB} (Proc Suppl.){\bf 83-84},521 (2000), {\bf 119},730(2003);  
{\NPA} {\bf 680},{151c}(2000),
hep-lat/0303024,0305010.
\bibitem{SHA} L. von Smekal, A. Hauck,  R. Alkofer, {Ann. Phys.}{\bf 267},1 (1998), hep-ph/9707327; 
\bibitem{lat03} H.Nakajima and S. Furui,Lattice2003 proceedings.
\bibitem{adelade} D.B. Leinweber, J.I. Skullerud, A.G. Williams and C. Parrinello, {\PRD}{\bf 60},{094507}{(1999)}; ibid {\PRD}{\bf 61},{079901}{(2000)}.
\bibitem{orsay1} D. Becirevic et al., {\PRD} {\bf 61},{114508}{(2000)}.
\bibitem{cuc} A. Cucchieri, {\NPB}{\bf508},{353}{(1997)}, hep-lat/9705005.
\bibitem{PMS} P.M.Stevenson, {\PRD}{\bf 23}, {2916}{(1981)}.
\bibitem{Gru} G. Grunberg, {\PRD}{\bf 29}, {2315}{(1984)}.
\bibitem{vAc} K. Van Acoleyen and H. Verschelde, {\PRD}{\bf 66},125012(2002),hep-ph/0203211.
\bibitem{HoMa} D.M. Howe and C.J. Maxwell, hep-ph/0204036 v2.
\bibitem{BroLu} S.J. Brodsky and H.J. Lu, {\PRD}{\bf 51},{3652}{(1995)}
\bibitem{BCLM} J.R.C.Bloch, A. Cucchieri, K.Langfeld and T.Mendes, hep-lat/0209040 v2. 
\bibitem{dudal} D.Dudal et al.,JHEP06(2003)003.
%,H. Vershelde, V.E.R. Lemes, M.S. Sarandy, S.P.Sorella, M.Picariello, A. Vicini and J.A. Gracey, JHEP06(2003)003.
\bibitem{KoShi} K-I. Kondo and T. Shinohara, {\PLB}{\bf 491},{263}{(2000)}.
\end{thebibliography}

%%%%%%%%%%%%%%%%%%%%%%%%%%%%%%%%%%%%%%%%%%%
%% Just a reminder that you may have to run bibtex
%% All of it up to \end{document} can be removed
%% if you don't like the warning.
%%%%%%%%%%%%%%%%%%%%%%%%%%%%%%%%%%%%%%%%%%%
\IfFileExists{\jobname.bbl}{}
 {\typeout{}
  \typeout{******************************************}
  \typeout{** Please run "bibtex \jobname" to optain}
  \typeout{** the bibliography and then re-run LaTeX}
  \typeout{** twice to fix the references!}
  \typeout{******************************************}
  \typeout{}
 }

\end{document}